\documentclass[11pt,a4paper]{article} 
\usepackage[dvips]{graphicx}
\usepackage{epstopdf}

\usepackage{amsmath}
\usepackage{epsfig}
\usepackage{t1enc}
\usepackage{amssymb}
\usepackage{latexsym}
\usepackage{bm}
\usepackage[hang]{caption}
\usepackage{ulem}

\usepackage{wrapfig}
 
\usepackage[extra]{tipa}

\title{{\bf Remarks on the Bojowald--Paily paper} \\ \textbf{\textit{
Deformed general relativity}}}

\author{Vladim\'ir Balek\footnote{e-mail
address: balek@fmph.uniba.sk}\ \ and Orchidea Maria Lecian
\footnote{e-mail address: omlecian@gmail.com}
\\
{\it Department of Theoretical Physics, Comenius University,
Bratislava, Slovakia}}

\begin{document}

\renewcommand{\figurename}{\small Fig.}

\maketitle \maketitle\abstract

{A couple of technicalities in the paper M. Bojowald, G. Paily,
Phys. Rev. D 87, 044044 (2013) are discussed. The explicit formula
given there for the function entering the modified
hypersurface-deformed algebra, which presumably originates in loop
quantum gravity, seems to be oversimplified, and the embedding of
deformed special relativity in deformed general relativity
proposed there for spherically symmetric models raises some
questions as well.}

\vskip 8mm An interesting problem in quantum gravity is to
determine how special relativity is deformed due to quantum
fluctuations of spacetime metric. Since deformations of special
relativity are restricted by symmetry requirements, predictions
obtained for them from a particular candidate for quantum theory
of gravitation can provide a test of its viability, or distinguish
between different approaches within it. Embedding deformed special
relativity in deformed general relativity within the framework of
loop quantum gravity (LQG) was examined in the paper by Bojowald
and Paily \cite{bopa}. The authors considered a deformation of
hypersurface-deformed algebra encoded in a single function
$\beta$, coming either from inverse triad corrections
\cite{{boho},{bore}} or from holonomy corrections
\cite{{rey},{cai}}, and obtained a Poincaré-type algebra in the
linear limit (the limit in which lapse and shift functions depend
linearly on spacetime coordinates). The resulting algebra turned
out to be incompatible with the much-studied $\kappa$-Poincaré
algebra, however, as was shown later \cite{ame}, this could be
cured by considering a more general representation of generators
of boosts and translations. The deformation of general relativity
by means of the function $\beta$ was used also in other contexts,
for example, when studying possible signature change at high
densities \cite{{miel},{bomie},{mietr}}. Here we discuss some
technical points in section II.C of \cite{bopa}, where the authors
consider linear limit of the deformed theory applied to
spherically symmetric models.

\vskip 2mm {\it Function $\beta$.}
In the modified Poisson bracket between Hamiltonian constraints
there appears a function $\beta$, which is, according to \cite{bopa},
in spherically symmetric models given by the function $F(K_\phi)$
replacing $K_\phi$
(angular component of extrinsic curvature) in the Hamiltonian. The
expression for $\beta$ is
\begin{equation}
\beta = \frac 12 \frac {d^2 F^2}{dK_\phi^2}.
\label{be0}
\end{equation}
For example, if we chose $F = \sin(\delta K_\phi)/\delta \equiv
F_\delta$, we obtain $\beta = \cos(2\delta K_\phi)$. Equation
(\ref{be0}) is surely true as long as the function $F$ does not
depend on $E^x$, $E^\phi$ (components of densitized triad).
However, the main result of \cite{bopa}, the formula
for the deformed Poisson bracket $\{B_x, P_0\}$, rests on the
assumption that the parameter $\delta$ in $F_\delta$ {\it does}
depend on $E^x$; namely, that it has the form $\delta \propto
|E^x|^{-1/2}$. Let us check whether equation (\ref{be0}), with
total derivative replaced by partial one, holds also for such
function. For that purpose, we will use a simplified version of
the theory explained in Appendix A.3 in \cite{rey}.

\vskip 2mm Hamiltonian constraint in a spherically symmetric model is
\begin{equation}
H[N] = - \frac 2G \int N |E^x|^{1/2} \Big[2K_x K_\phi + \frac
{E^\phi}{|E^x|}K_\phi^2 + 2\Gamma_\phi' + \frac {E^\phi}{|E^x|}
(1 - \Gamma_\phi^2)\Big]dx,
\label{H0}
\end{equation}
where $K_x$ is radial component of extrinsic curvature and
$\Gamma_\phi$ is angular component of spin connection,
$\Gamma_\phi = - {E^x}'/(2E^\phi)$. (All quantities depend on
radial coordinate $x$ only.) By using Poisson brackets between
components of extrinsic curvature and densitized triad,
$$\{K_x (x), E^x (y)\} = \{K_\phi (x), 2E^\phi (y)\} = 2G \delta(x -
y),$$ we find that the Poisson bracket between Hamiltonian
constraints can be written as
\begin{equation}
\{H[M], H[N]\} = \frac 1G \int (MN' - NM') \frac {|E^x|}{E^\phi}
(K_\phi' + \Gamma_\phi K_x) dx.
\label{Pb0}
\end{equation}
On the right hand side there appears the diffeomorphism constraint
$D[\vec {\cal N}]$, with the shift vector $\vec {\cal N}$ given in terms
of lapse functions $M$, $N$ as $\vec {\cal N} = {\cal N}^x \partial_x$,
${\cal N}_x = MN' - NM'$.

\vskip 2mm Now, put instead of $K_\phi$ into the first and second
term in $H[N]$ some functions $F_2$ and $F_1$, both depending on
$K_\phi$, $E^x$ and $E^\phi$. Thus, replace
$$2K_x K_\phi + \frac {E^\phi}{|E^x|} K_\phi^2\ \to\ 2K_x F_2 +
\frac {E^\phi}{|E^x|} F_1^2.$$ This yields Poisson bracket
\begin{equation}
\{H[M], H[N]\} = \frac 1G \int (MN' - NM') \Big[\frac
{|E^x|}{E^\phi} \Big(F_2' + \Gamma_\phi K_x \frac {\partial
F_2}{\partial K_\phi}\Big) + \Gamma_\phi \Big(\frac 12 \frac
{\partial F_1^2}{\partial K_\phi} - F_2\Big) \Big]dx. \label{Pb}
\end{equation}
We wish that the right hand side can be written as $D[\beta \vec
{\cal N}]$ with some function $\beta$ depending on $K_\phi$, $E^x$ and
$E^\phi$. Clearly, in case both $F_1$ and $F_2$ depend only on
$K_\phi$, we achieve this by putting
\begin{equation}
F_2 = \frac 12 \frac {dF_1^2}{dK_\phi}, \label{cond0}
\end{equation}
since then the last two terms in the square brackets cancel and
the right hand side acquires the desired form with
\begin{equation}
\beta = \frac {dF_2}{dK_\phi}. \label{be}
\end{equation}
After inserting here for $F_2$, we obtain equation (\ref{be0})
with $F = F_1$.

\vskip 2mm Suppose now that the functions $F_1$, $F_2$ depend also
on $E^x$, $E^\phi$. Then, in order to achieve the
desired form of the Poisson bracket, two requirements must be
satisfied: first, the function $F_2$ must not depend on $E^\phi$
(and, as a result, the function $F_1^2$ may contain $E^\phi$
only additively); and second, equation
determining $F_2$ in terms of $F_1$ must be modified to
\begin{equation}
F_2 + 2 E^x \frac {\partial F_2}{\partial E^x} = \frac 12 \frac
{\partial F_1^2}{\partial K_\phi}. \label{cond}
\end{equation}
For the function $\beta$ we have again equation (\ref{be}), we
just have to replace the total derivative by the partial one,
\begin{equation}
\beta = \frac {\partial F_2}{\partial K_\phi}. \label{be1}
\end{equation}
The extra term on the left hand side of (\ref{cond}) makes the
expression of $\beta$ in terms of $F_1$ more complicated: instead
of equation (\ref{be0}) with $F = F_1$ we have
\begin{equation}
\beta = |E^x|^{-1/2} \int \limits_0^{|E^x|^{1/2}} \frac 12 \frac
{\partial^2 \bar F_1^2}{\partial K_\phi^2} d(|\bar E^x|^{1/2}).
\label{bmod}
\end{equation}
where $\bar F_1 = F_1(K_\phi, \bar E^x)$. (We have chosen the limits
of the integral so that $\beta \to 1$ in the limit $|E^x| \to \infty$,
when presumably $F_1 \to K_\phi$.) In particular, if $F_1 =
F_\delta$ with $\delta \propto |E^x|^{-1/2}$, the expression
for $\beta$ is, rather that $\beta = \cos u$,
\begin{equation}
\beta = \int \limits_0^1 \cos (u\xi^{-1}) d\xi,
\label{bmod1}
\end{equation}
where $u = 2\delta K_\phi$. This can be written as
$\beta = \cos u - (\pi/2) |u| + \mbox{Si}(u)u$,
where Si is sine integral, Si$(u) = \displaystyle
\int \limits_0^u \frac {\sin \bar u}{\bar u} d\bar u$.
The asymptotics of $\beta$ are $\beta \doteq 1 - (\pi/2) |u| +
(1/2) u^2$ for $|u| \ll 1$ and $\beta \doteq -\sin u/u$
for $|u| \gg 1$; thus, $\beta$ falls linearly rather than
quadratically if $u$ is close to 0, and undergoes
damped oscillations rather than keeping its amplitude constant
if $u$ is far from 0.

\vskip 2mm {\it Radial momentum.} According to \cite{bopa}, the
momentum stored in the region $\Sigma$ of the hypersurface of
constant time, projected onto a given vector $v^a$, is
\begin{equation}
P = 2\int \limits_{\partial \Sigma} v_a (p^{ab} r_b - \bar p^{ab}
\bar r_b) d^2 z,
\label{dfP}
\end{equation}
where $p^{ab}$ is the momentum canonically conjugated with $h_{ab}$
(metric tensor on $\Sigma$), $r^a$ is normal to $\partial \Sigma$,
$\bar p^{ab}$ and $\bar r^a$ is momentum and normal computed for
reference metric, and $(z^1, z^2)$ are coordinates on $\partial
\Sigma$. For a spherically symmetric metric this yields, according
to \cite{bopa}, the formula for radial component of momentum
\begin{equation}
P_x = \frac {8\pi h_{xx} p^{xx}}{E^\phi |E^x|^{1/2}}.
\label{Px}
\end{equation}
After inserting here $h_{xx} = (E^\phi)^2/|E^x|$ and $16\pi G p^{xx}
= h^{1/2} (K^{xx} - h^{xx} K) = - 2h^{1/2} h^{xx} K_\theta^\theta
= - 2h^{1/2}$ {\scriptsize $\times$ \linebreak $\times$} $h^{xx}
(h^{\theta \theta})^{1/2} K_\phi = -2(|E^x|/E^\phi)
K_\phi$ (the function $p^{xx}$ in equation (\ref{Px}) is
supposed to be stripped of the factor $\sin \theta$, therefore we
suppressed the factor also in the function $h^{1/2}$), we can
write
\begin{equation}
P_x = - \frac 1G \frac {K_\phi}{|E^x|^{1/2}}.
\label{Px1}
\end{equation}
This is used later -- albeit in combination with expression (\ref{be0}) for
$\beta$, which is apparently flawed -- in the derivation of the deformed
Poisson bracket $\{B_x, P_0\}$.

\vskip 2mm The problem with the expression for $P_x$ is the
denominator of the fraction on the right hand side of equation
(\ref{Px}). It is unclear where this denominator, which equals
$h^{1/2}$, comes from. To see that, let us look closer at the
definition of $P$.

\vskip 2mm The object $v_a$ appearing in (\ref{dfP}) is identified
in \cite{bopa} by an expression for $v^a$ which
clearly represents a vector, hence $v_a$ is a co-vector; and the
object $r_a$ is called ``co-normal'' in \cite{bopa}, which
suggests that it is a co-vector too. However, if this was the case,
the expression for $P$ could not be correct since it would not be invariant
with respect to diffeomorphisms on $\Sigma$. In \cite{bojo}, a
different expression for $P$ is given (although it is not called
so) in which $v^a$ is identified with shift vector, $v^a = N^a$.
The expression contains an additional factor $1/N$ in the
integral, with $N$ previously identified as lapse function,
$$P = 2\int \limits_{\partial \Sigma} N_a (p^{ab} r_b - \bar p^{ab} \bar
r_b) \frac {d^2 z}N.$$ This cannot be correct either, since $N$ is
scalar and its presence in the integral does not change
transformation properties of $P$. In order to obtain a meaningful
expression, we must replace $N \to n = dl/dx$, where $x$ is
transversal coordinate (a coordinate varying in the direction
orthogonal to $\partial \Sigma$) and $l$ is distance measured
along $x$. Clearly, $n$ plays the same role in $2 + 1$
decomposition as $N$ plays in $3 + 1$ decomposition. Thus, instead
of (\ref{dfP}) we have
\begin{equation}
P = 2\int \limits_{\partial \Sigma} v_a (p^{ab} r_b - \bar p^{ab} \bar
r_b) \frac {d^2 z}n.
\label{dfP1}
\end{equation}
Note that one can easily check that this expression is right by computing $P$
as a surface term in diffeomorphism constraint.

\vskip 2mm In a spherically symmetric metric, in which we chose
$\Sigma$ as a ball with the center in the center of symmetry, we
have $r_a = h_{xx}^{1/2} \delta_a^x$ and $n = h_{xx}^{1/2}$, hence
$r_a/n = \delta_a^x$; furthermore, the vector $v^a$ is identified
as $v^a = (\partial/\partial x)^a = \delta_x^a$ in \cite{bopa},
hence $v_a = h_{ax} = h_{xx} \delta_a^x$. By inserting this into
equation (\ref{dfP1}) and using $\bar p^{xx} = 0$, we arrive at an
expression for $P_x$ different from (\ref{Px1}),
\begin{equation}
P_x = 8\pi h_{xx} p^{xx} = - \frac 1G E^\phi K_\phi.
\label{Pmod}
\end{equation}

\vskip 2mm If the previous considerations are true and
expression (\ref{Px1}) for $P_x$ is indeed wrong, using it seems
to be more serious problem of the theory than just using an
oversimplified expression for the function $\beta$. For the
reasoning in \cite{bopa} it is crucial that the expression for
$P_x$ contains the same combination of the variables $K_\phi$,
$E^x$ and $E^\phi$ as appears in $\beta$; however, as we have
seen, whatever the combination is, it cannot include the variable
$E^\phi$. This remains true also for more general
modification of the Hamiltonian constraint than considered here,
in which one uses, in addition to the functions $F_1$ and $F_2$,
four more functions of the variables $E^x$ and $E^\phi$ denoted by
$\alpha$, $\bar \alpha$, $\alpha_\Gamma$ and $\bar \alpha_\Gamma$
\cite{rey}. Thus, if the expression (\ref{Px1}) has indeed to be
replaced by the expression (\ref{Pmod}), the deformed bracket
$\{B_x, P_0\}$ cannot be given by the simple formula from
\cite{bopa}. Perhaps it can be still computed within the framework
of \cite{bopa}, but its exact form remains to be found.

\vskip 2mm {\it Undeformed bracket $\{B_x, P_0\}$.} The main point
of \cite{bopa} is that we can test LQG by obtaining a hint from it
about how the bracket $\{B_x, P_0\}$ is deformed due to quantum
gravity effects. In order to do so we need to know how the
undeformed bracket looks like. According to \cite{bopa}, it holds
\begin{equation}
\{B_x, P_0\} = P_x, \label{BP0}
\end{equation}
see equation (18) in section II.C in \cite{bopa}, in which one has
to put $\lambda = 0$. Thus, the equation relating, via Poisson
bracket, radial components of the boost generator $\bf B$ and
the momentum $\bf P$, is according to \cite{bopa} the same as the
equation relating their Cartesian components.

\vskip 2mm Equation (\ref{BP0}) refers to {\it special}
relativity; thus, the quantities $P_0$, $P_x$ and $B_x$ appearing
there are presumably integral characteristics of a classical
system in flat spacetime. However, if this is the case, we must
add an extra term proportional to $\dot P_x$ to the right hand
side of equation (\ref{BP0}),
\begin{equation}
\{B_x, P_0\} = P_x - \dot P_x t. \label{BP}
\end{equation}
This is most easily seen if we express $B_x$ in terms of $P_0$ and
$P_x$ and use the fact that for any given dynamical quantity $f$ it
holds $\{f, P_0\} = - \{f, H\} = - \dot f$. Denote the energy density
by $\epsilon$ and the total energy of the system by $E$, $E =
\displaystyle \int \epsilon d^3 x$. Radial component of the
boost generator can be written as $B_x = XP_0 - x_0 P_x = -EX + P_x t$,
where $X$ is the ``radial center-of-mass coordinate'',
$$X = \frac 1E \int x \epsilon d^3 x.$$
With this expression for $B_x$, the bracket becomes
$$\{B_x, P_0\} = - E\{X, P_0\} + \{P_x, P_0\} t =
E\dot X - \dot P_x t.$$
Furthermore, the energy density satisfies the continuity
equation $\dot \epsilon + \nabla \cdot {\bf p} = 0$, where $\bf p$ is
the momentum density, therefore the first term on the right hand side
can be written as
$$E\dot X = \int x \dot \epsilon d^3 x = - \int x \nabla \cdot {\bf p} d^3 x
= \int \nabla x \cdot {\bf p} d^3 x = \int p_x d^3 x = P_x.$$
(We have skipped the surface integral since we assume that the momentum
density falls off rapidly enough as $x$ goes to infinity.)
As a result, we find that the bracket is given by equation (\ref{BP}).

\vskip 2mm The time derivative of the Cartesian components
of momentum is of course zero, however, for other components
such as the radial one this is no longer true. For example, for
a system of nonrelativistic particles with radius vectors
${\bf x}_\alpha$ and interaction potential $v({\bf x}_\alpha)$
we have
\begin{equation}
\dot P_x = \sum \frac {\partial v}{\partial x_{\alpha \beta}}
\frac {x_\alpha - {\bf n}_\alpha \cdot {\bf n}_\beta x_\beta}
{x_{\alpha \beta}}, \label{Pdot}
\end{equation}
where ${\bf n}_\alpha =
{\bf x}_\alpha/x_\alpha$ and $x_{\alpha \beta} = |{\bf x}_\alpha
- {\bf x}_\beta|$; and for scalar field $\phi$ with potential
$V(\phi)$ we have
\begin{equation}
\dot P_x = \int \frac {2l}x d^3 x, \quad l = \frac 12
\pi^2 - \frac 12 (\partial_x \phi)^2 - V, \label{Pdot1}
\end{equation}
where $\pi$ is the momentum canonically conjugated with $\phi$, $\pi =
\dot \phi$. (The function $l$ coincides with Lagrangian density in
case the field is spherically symmetric). As we can see, the
second term in (\ref{BP}) cannot be ignored, so that a question
arises whether LQG can make any predictions concerning its deformation,
as it presumably does with respect to the first term.

\end{document}